\documentclass[twocolumn, tighten, times]{aastex62}
\usepackage{graphicx}
\usepackage{hyperref}
\usepackage{amsmath}
\usepackage{natbib}
\usepackage{soul}
\setstcolor{red}

\newcommand{\Planck}{{\it Planck}}

\graphicspath{{./}{figures/}}

\received{xxx}
\revised{yyy}
\accepted{zzz}
\submitjournal{ApJ}

\begin{document}
\title{Measuring the Hubble constant from the cooling of the CMB monopole}
\author[0000-0001-8071-6735]{Maximilian~H.~Abitbol}
\email{maximilian.abitbol@physics.ox.ac.uk}
\affiliation{University of Oxford, Department of Physics, Denys Wilkinson Building, Keble Road, Oxford OX1 3RH, UK}
\author{J.~Colin Hill}
\email{jch2200@columbia.edu}
\affiliation{Department of Physics, Columbia University, New York, NY, USA 10027}
\affiliation{Center for Computational Astrophysics, Flatiron Institute, New York, NY, USA 10010}
\affiliation{School of Natural Sciences, Institute for Advanced Study, Princeton, NJ, USA 08540}
\author{Jens Chluba}
\email{jens.chluba@manchester.ac.uk}
\affiliation{Jodrell Bank Centre for Astrophysics, School of Physics and Astronomy, Alan Turing Building, The University of Manchester, Manchester, M13 9PL, UK}

%%%%%%%%%%%%%%%%%%%%%%%%%%%%%%%%%%%%%%%%%%%%%%%%%%%%%%%%%%%%%%%%%%%%%%%%%%%%%%%%%%%%%%%%
\begin{abstract}
The cosmic microwave background (CMB) monopole temperature evolves with the inverse of the cosmological scale factor, independent of many cosmological assumptions. With sufficient sensitivity, real-time cosmological observations could thus be used to measure the local expansion rate of the Universe using the cooling of the CMB. We forecast how well a CMB spectrometer could determine the Hubble constant via this method. The primary challenge of such a mission lies in the separation of Galactic and extra-Galactic foreground signals from the CMB at extremely high precision. However, overcoming these obstacles could potentially provide an independent, highly robust method to shed light on the current low-/high-$z$ Hubble tension. An experiment with 3000 linearly spaced bins between 5~GHz and 3~THz with a sensitivity of 1~$\mathrm{mJy\sqrt{yr}~sr^{-1}}$ per bin, could measure $H_0$ to 3\% over a 10 year mission, given current foreground complexity. This sensitivity would also enable high-precision measurements of the expected $\Lambda$CDM spectral distortions, but remains futuristic at this stage. 
\end{abstract}


%%%%%%%%%%%%%%%%%%%%%%%%%%%%%%%%%%%%%%%%%%%%%%%%%%%%%%%%%%%%%%%%%%%%%%%%%%%%%%%%%%%%%%%%
\section{Introduction} 
\label{sec:intro}
Measurements of the expansion of the Universe serve as a cornerstone of modern cosmology. The two most precise methods to infer the current expansion rate, $H_0$, are i) ``direct'' observations using Type Ia supernovae as standard candles, calibrated using Cepheid variables or other local objects with precise distance measurements, and ii) ``indirect'' determinations using the standard-ruler sound horizon measured by cosmic microwave background (CMB) anisotropy observations.  The latter method can also be employed without CMB data by using baryon acoustic oscillation, Big Bang nucleosynthesis, and weak lensing data~\citep[e.g.,][]{des2018h0}. These methods infer the Hubble constant via physics at opposite ends of cosmic history, providing an important consistency check on $\Lambda$CDM cosmology. 
Current results from the S$\mathrm{H_0}$ES team, using Cepheids in the Large Magellanic Cloud to calibrate the distance ladder, yield $H_0 = 74.03 \pm 1.42$~km~s$^{-1}$~Mpc$^{-1}$~\citep{riess2019}. \Planck\ observations of the CMB anisotropies result in $H_0=67.4 \pm 0.50$~km~s$^{-1}$~Mpc$^{-1}$, indicating a $4.4\sigma$ tension with S$\mathrm{H_0}$ES~\citep{planck2018cosmology}.

The discrepancy in $H_0$ derived by these two methods has garnered much attention and led to the development of a variety of cosmological and experimental explanations \citep[see][for some examples]{knox2019_hubblehunters}. Several additional pathways have been used to measure $H_0$ including the use of: different observables to build the distance ladder~\citep{freedman2019_cchp, huang2019mira}; strong gravitational lensing of quasars~\citep{wong2019holicow,strides2019lens}; gravitational waves as standard sirens~\citep{ligo2019hubble}; and many others, as well as reanalyses of these data with new methods~\citep{reid2013megamaser, jimenez2019stellar, kozmanyan2019xray}. In this paper we forecast the use of real-time cosmology to measure $H_0$ from the Hubble cooling of the CMB monopole temperature. Similar real-time cosmological methods have been proposed using redshift drift~\citep[e.g. using the Sandage-Loeb test][]{sandage1962,loeb1198}.

The time evolution of the temperature of the CMB monopole provides a local measurement of the universal expansion rate, without requiring use of a distance ladder. Additionally, the theoretical assumptions for this type of measurement are extremely minimal and allow for a direct measurement of the time evolution of the scale factor, without assuming a particular cosmology. The evolution of the CMB monopole temperature with the scale factor is a well-known result: $T_{\rm CMB} \propto a^{-1}$. This relation cannot be easily modified without introducing noticeable CMB spectral distortions \citep[e.g.,][]{Chluba2014TRR}.  The only underlying assumptions involved in this relation are:
\begin{itemize}
    \item Photons are massless;
    \item The CMB is thermal radiation (i.e., if the photon distribution function was far from Planckian, then $T_{\rm CMB}$ would not be a meaningful quantity);
    \item The first law of thermodynamics is valid;
    \item The expansion of space is isotropic.
\end{itemize}
All of these assumptions have already been validated experimentally at extremely high precision.

Thus, the quantities $T_{\rm CMB}$ and $a$ can be mapped exactly onto one another. The evolution of the temperature of the CMB with time then provides an unambiguous, direct measurement of $H_0 \equiv \left(\frac{\dot{a}}{a}\right)|_{t=t_0}$. The primary obstacle comes in the form of Galactic and extra-Galactic foregrounds that obscure our view of the CMB~\citep{abitbol2017}. Experimental challenges arise in the sensitivity, stability, and calibration requirements necessary for such an observation. In this paper we present the method and forecast noise levels required to measure the Hubble constant using the cooling of the CMB.

The time evolution of the monopole as a probe for $H_0$ has been proposed previously~\citep{zibin2007cmbevolution}. We take the additional step of propagating the uncertainties on $T_{\rm CMB}$ to $H_0$ and present example noise curves to produce such a measurement, including a detailed treatment of foregrounds. Studies of the time evolution of the CMB angular power spectrum and forecasts for observations have also been conducted~\citep{Lange07thetime,zibin2007cmbevolution,moss2008cmbevolution}. In particular, \citet{lange_page2007} found that the change in the power spectrum could in principle be detected over the course of a century. 

%%%%%%%%%%%%%%%%%%%%%%%%%%%%%%%%%%%%%%%%%%%%%%%%%%%%%%%%%%%%%%%%%%%%%%%%%%%%%%%%%%%%%%%%
\section{Method}
\label{sec:method}

\subsection{Hubble cooling of $T_{\rm CMB}$}
The monopole temperature of the CMB, hereafter denoted $T \equiv T_{\mathrm{CMB}}$, decreases in time with the scale factor as
%%%%%%%%%%%%%%%%%%%%%%%%%%%%%%%%%%%%%%%%%%%%%%%%%%%%%%%%%%%%%%%%%%%%%%%%%%%%%%%%%%%%%%%%
\begin{equation}
    T(t) = T_0/a(t) \,.
\end{equation}
%%%%%%%%%%%%%%%%%%%%%%%%%%%%%%%%%%%%%%%%%%%%%%%%%%%%%%%%%%%%%%%%%%%%%%%%%%%%%%%%%%%%%%%%
The time derivative of $T$ is then $\dot{T}(t) = - H(t) T(t)$. This can be evaluated to first order, taking $H_0=70~\mathrm{km~s^{-1}~Mpc^{-1}}$ $= 7.2\times 10^{-11}~\mathrm{yr^{-1}}$ as an example and $T_0=2.725$~K,\footnote{The forecast for $\sigma_{H_0}$ depends on the true value of the current CMB temperature $T_0$ but the change in $\sigma_{H_0}$ is small given the current sub-mK constraints on $T_0$ from COBE/FIRAS \citep{Fixsen1996,Fixsen2011}.}
%%%%%%%%%%%%%%%%%%%%%%%%%%%%%%%%%%%%%%%%%%%%%%%%%%%%%%%%%%%%%%%%%%%%%%%%%%%%%%%%%%%%%%%%
\begin{equation}
    \dot{T}_0 = H_0 T_0 = -0.20~\mathrm{nK\, yr^{-1}} \,. 
\end{equation}
%%%%%%%%%%%%%%%%%%%%%%%%%%%%%%%%%%%%%%%%%%%%%%%%%%%%%%%%%%%%%%%%%%%%%%%%%%%%%%%%%%%%%%%%
Therefore the temperature of the CMB decreases by 2~$\mathrm{nK}$ over a 10 year period. One could thus in principle measure the real-time cooling of the CMB due to the expansion of space and from this infer the value of the Hubble constant.

\subsection{Requirements to measure $H_0$}
The simplest method to measure the Hubble constant from the monopole temperature would amount to fitting for the slope of $T$ as it decreases in time. The model is linear in time with a slope of $H_0 T_0$,
%%%%%%%%%%%%%%%%%%%%%%%%%%%%%%%%%%%%%%%%%%%%%%%%%%%%%%%%%%%%%%%%%%%%%%%%%%%%%%%%%%%%%%%%
\begin{equation}
T(t) = T_0 - H_0 T_0 t \,.
\label{eqn:Tt}
\end{equation}
%%%%%%%%%%%%%%%%%%%%%%%%%%%%%%%%%%%%%%%%%%%%%%%%%%%%%%%%%%%%%%%%%%%%%%%%%%%%%%%%%%%%%%%%
We have in mind a future version of the COBE/FIRAS experiment, where the data can be reduced in discrete time intervals to evaluate the monopole temperature of the CMB repeatedly over the mission duration. For example, the mission could take data continuously for 10 years, potentially mapping out large portions of the sky, and the monopole CMB temperature would be inferred from these data every year. Given the expected level of cooling, we can then propagate uncertainties from $T(t)$ to $H_0$ and determine what sensitivity to $T(t)$ is required to measure $H_0$ with a given precision.

We make some simplifying assumptions that will allow us to analytically calculate the uncertainty on $H_0$ using an ordinary least-squares estimator. First, we assume that $T(t)$ is measured with (foreground-marginalized) uncertainty $\sigma_T$ in discrete uniform time intervals with spacing $\Delta t$, such that $t=k\Delta t$, where $k=1,2...,N$. The mission duration is thus $t_{\rm total}=N\Delta t$ with $N$ data points. Second, the uncertainties on $T(t)$ are time-independent, parameter-independent, and Gaussian uncorrelated.\footnote{This is not entirely true as the CMB photon noise is one of the dominant noise sources and therefore the noise depends on $T_{\rm CMB}$.  However, the change in the CMB photon noise over these time scales is negligible.} The foreground-marginalized uncertainty $\sigma_T$ can then be written in terms of an effective sensitivity $s_T$ of the experiment such that, 
%%%%%%%%%%%%%%%%%%%%%%%%%%%%%%%%%%%%%%%%%%%%%%%%%%%%%%%%%%%%%%%%%%%%%%%%%%%%%%%%%%%%%%%%
\begin{equation}
\sigma_T = \frac{s_T}{\sqrt{\Delta t}} \,,
\end{equation}
%%%%%%%%%%%%%%%%%%%%%%%%%%%%%%%%%%%%%%%%%%%%%%%%%%%%%%%%%%%%%%%%%%%%%%%%%%%%%%%%%%%%%%%%
where $s_T$ is understood as the effective sensitivity to $T$ of the experiment after data reduction and foreground marginalization. The SI units for $\sigma_T$ and $s_T$ are $\mathrm{K}$ and $\mathrm{K\sqrt{s}}$, respectively. With these assumptions, we can use a linear least-squares regression to propagate the uncertainties from $T(t)$ to $H_0$ and solve this in the limit that $H_0\Delta t\ll 1$ and $\sigma_{T_0} \ll T_0$, 
%%%%%%%%%%%%%%%%%%%%%%%%%%%%%%%%%%%%%%%%%%%%%%%%%%%%%%%%%%%%%%%%%%%%%%%%%%%%%%%%%%%%%%%%
\begin{equation}
  \sigma_{H_0} = \frac{s_T}{T_0 t^{3/2}_{\rm total}} \sqrt{\frac{12 N^2}{N^2-1}} \approx \sqrt{12}\frac{s_T}{T_0 t^{3/2}_{\rm total}} \,.
  \label{eqn:sigmaH0}
\end{equation}
%%%%%%%%%%%%%%%%%%%%%%%%%%%%%%%%%%%%%%%%%%%%%%%%%%%%%%%%%%%%%%%%%%%%%%%%%%%%%%%%%%%%%%%%

\begin{figure*}[ht]
  \includegraphics[width=\textwidth]{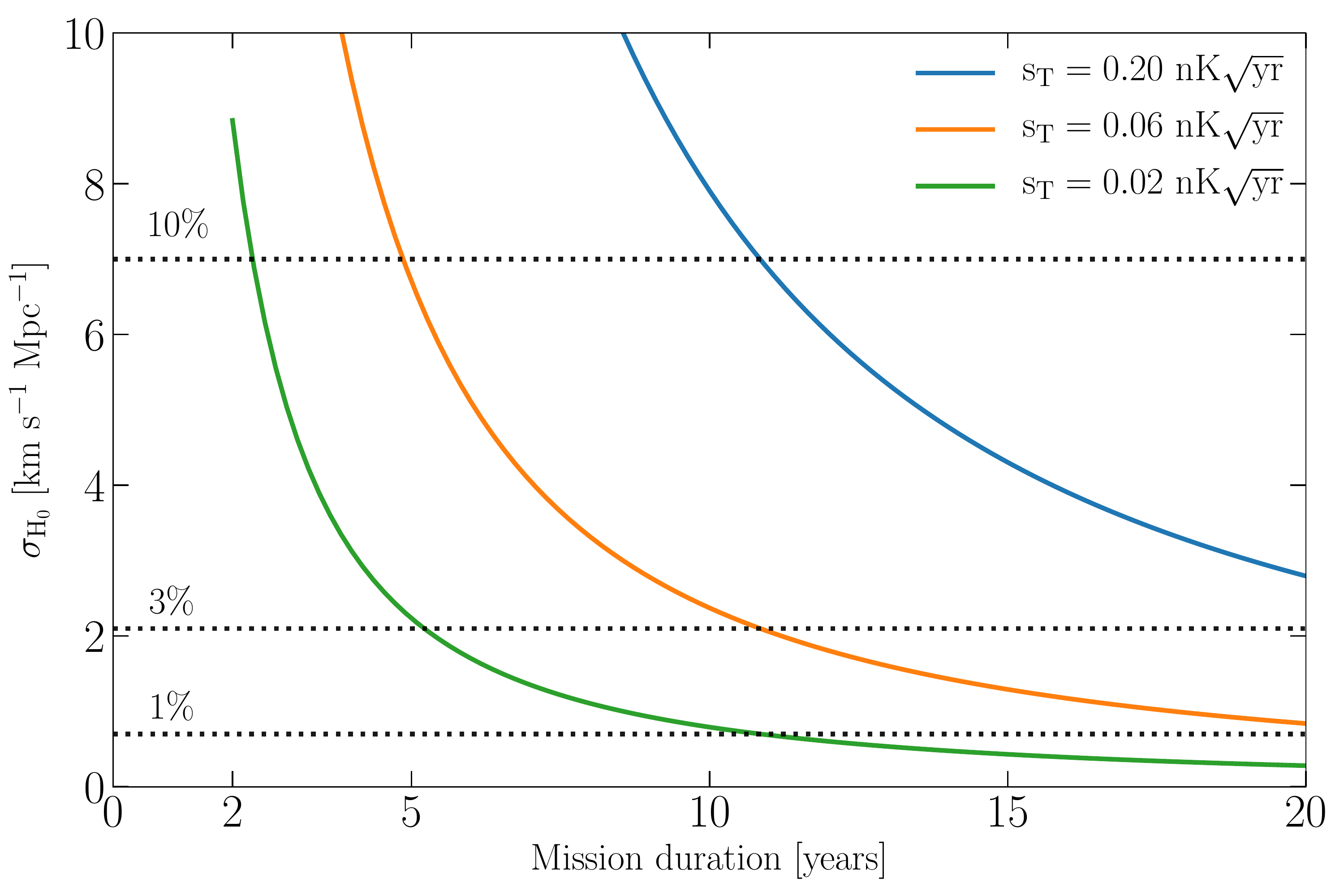}
    \caption{Hubble constant uncertainty as a function of mission duration for different effective sensitivities to $T$. The scaling of $\sigma_{H_0}$ depends on mission duration to the $-3/2$, as the noise averages down with the square root of time and the signal increases linearly in time. An experiment with effective monopole temperature sensitivity $s_T=0.06~\mathrm{nK \, \sqrt{yr}}$ could measure $H_0$ to $3\%$ in about 11 years (see Equation~\ref{eqn:sigmaH0}). }
    \label{fig:sensitivity}
\end{figure*}

Figure~\ref{fig:sensitivity} shows the uncertainty on $H_0$ as a function of mission duration and effective sensitivity. To minimize the uncertainty on $H_0$ requires minimizing $s_T$ and maximizing $t_{\rm total}$, as expected. The uncertainty on $H_0$ has a linear dependence on $s_T$ and a $3/2$ scaling with the mission duration. The power of $3/2$ comes from the signal increasing linearly with time and the noise averaging down as the square root of time. The term in the square root is nearly independent of $N$ for $N>2$ and so the choice of $N$ (or how to divide up the data for a mission with fixed duration) is not too important.

We can then calculate the effective sensitivity $s_T$ to measure $H_0$ at a given precision $X$, where $X = H_0 / \sigma_{H_0}$ is the detection significance of $H_0$:
%%%%%%%%%%%%%%%%%%%%%%%%%%%%%%%%%%%%%%%%%%%%%%%%%%%%%%%%%%%%%%%%%%%%%%%%%%%%%%%%%%%%%%%%
\begin{equation}
    s_T = \frac{0.056}{X} \left(\frac{t_{\rm total}}{1~\mathrm{year}}\right)^{3/2}~\mathrm{[nK \sqrt{yr}]}
\end{equation}
%%%%%%%%%%%%%%%%%%%%%%%%%%%%%%%%%%%%%%%%%%%%%%%%%%%%%%%%%%%%%%%%%%%%%%%%%%%%%%%%%%%%%%%%
For a 10-year mission this gives 
%%%%%%%%%%%%%%%%%%%%%%%%%%%%%%%%%%%%%%%%%%%%%%%%%%%%%%%%%%%%%%%%%%%%%%%%%%%%%%%%%%%%%%%%
\begin{equation}
    s_T = \frac{1.8}{X}~\mathrm{nK \sqrt{yr}} \,.
\end{equation}
%%%%%%%%%%%%%%%%%%%%%%%%%%%%%%%%%%%%%%%%%%%%%%%%%%%%%%%%%%%%%%%%%%%%%%%%%%%%%%%%%%%%%%%%
To be comparable with the state-of-the-art measurements of $H_0$ with $3\%$ uncertainties or less requires $X\gtrsim 30$, thus an effective experiment sensitivity of $s_T \lesssim 0.059~\mathrm{nK \sqrt{yr}}$ for a 10-year mission.  We will compare this benchmark noise level to recently proposed and upcoming experiments in the next section.

%%%%%%%%%%%%%%%%%%%%%%%%%%%%%%%%%%%%%%%%%%%%%%%%%%%%%%%%%%%%%%%%%%%%%%%%%%%%%%%%%%%%%%%%
\section{Results}
\label{sec:results}
\subsection{Foreground assumptions}
\label{sec:fgs}

\begin{figure*}[ht]
  \includegraphics[width=\textwidth]{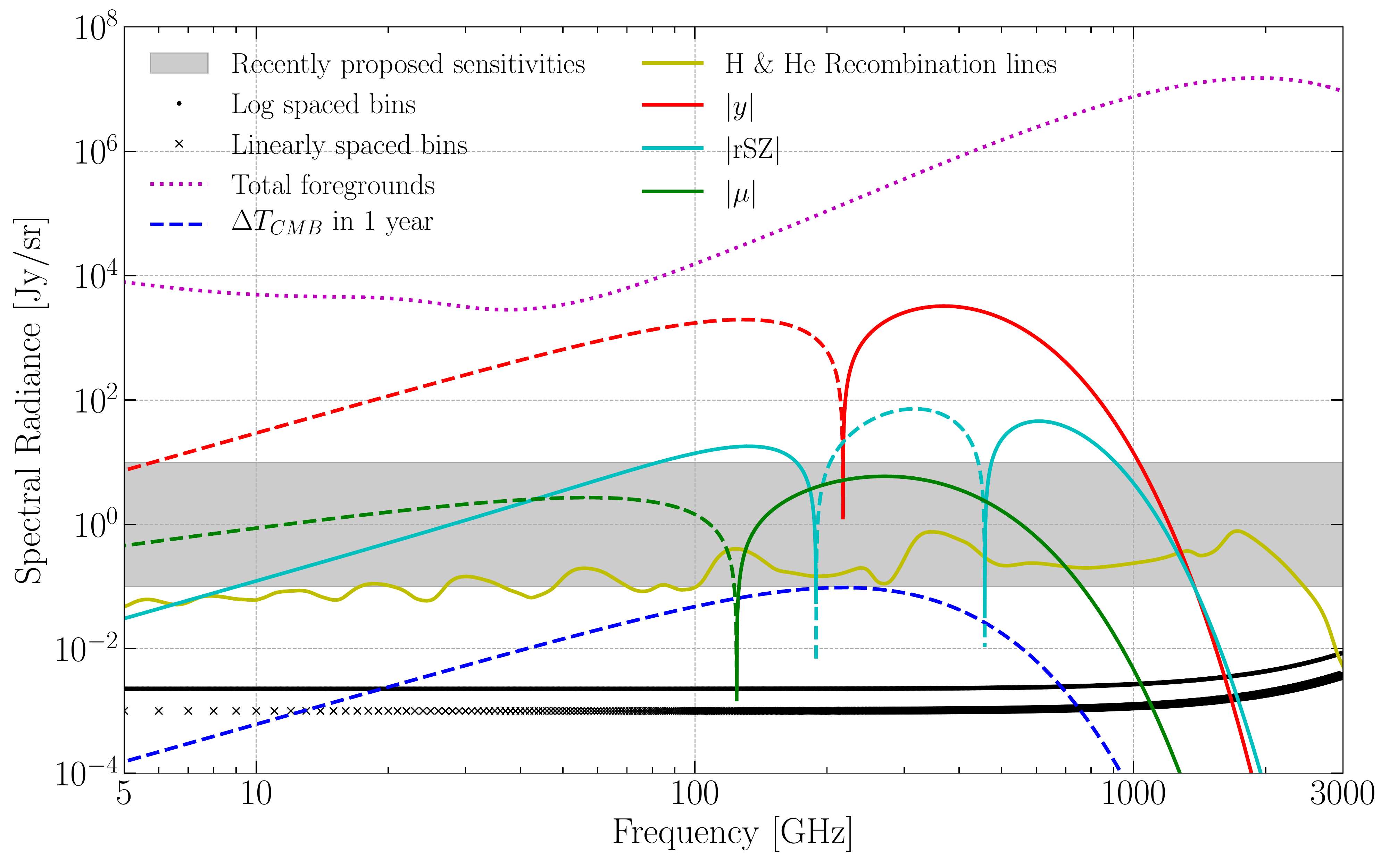}
  \caption{Sky signals and sensitivities. In dotted purple is a representative model for the total foregrounds~\citep{abitbol2017}. CMB spectral distortion signals are shown in the solid (and partially dashed) curves, with the Compton-$y$ distortion in red, relativistic correction to $y$ in cyan, chemical potential $\mu$ distortion in green and hydrogen and helium recombination lines in yellow. The change in the CMB monopole after one year is plotted in blue. The signals are plotted with respect to a nominal CMB blackbody at 2.725~K, with the negative parts shown as dashed curves. Two example sensitivity curves are plotted in black, one with linearly spaced bins and one with log spaced bins. Both example sensitivities are designed to give a $3\%$ measurement of $H_0$ with a 10 year mission duration (the sensitivity per year is shown). The shaded region shows the approximate sensitivity range from recently proposed spectral distortion missions achievable with current technology.}
    \label{fig:signals}
\end{figure*}

One of the main limitations in making this measurement is the problem of separating out the CMB signal from a variety of microwave foreground signals. We use the Fisher matrix methodology described in~\citet{abitbol2017} to forecast uncertainties on parameters given an experimental noise model and a parametric model for the sky signals. The relevant CMB signals include the $y$-distortion, relativistic correction to $y$, $\mu$-distortion, and hydrogen and helium recombination lines. The foregrounds included in the model are:
\begin{itemize}
    \item Galactic synchrotron (3 parameters)
    \item Galactic free-free (1 parameter)
    \item Galactic anomalous microwave emission (1 parameter)
    \item Galactic thermal dust (3 parameters)
    \item integrated extra-Galactic CO line emission (1 parameter)
    \item cosmic infrared background (CIB) (3 parameters)
\end{itemize}
 for a total of 12 foreground parameters and 5 CMB distortion parameters ~\citep[see][for a detailed description of the model parameterization]{abitbol2017}. At this level of precision, the foregrounds might be much more complex and more general modeling \citep[e.g., as in][to account for spatial averaging effects]{Chluba2017} would be required. Additional foregrounds such as atomic and molecular lines in our Galaxy or zodiacal emission will contribute as well, but are ignored in this forecast, assuming spatial information can be used to mask or constrain them.

%%%%%%%%%%%%%%%%%%%%%%%%%%%%%%%%%%%%%%%%%%%%%%%%%%%%%%%%%%%%%%%%%%%%%%%%%%%%%%%%%%%%%%%%
\subsection{Sensitivity calculation}
\label{sec:sens}
In order to fit for all the foreground and spectral distortion parameters, we require an experiment with broad frequency coverage and high sensitivity. As seen in Figure~\ref{fig:signals}, the time-evolution signal is six orders of magnitude below the foregrounds and also below the other spectral distortion signals. As an example of a currently feasible mission, we begin by taking the sensitivity from the previously proposed, absolutely-calibrated Fourier transform spectrometer (FTS), PIXIE (Primordial Inflation Explorer). The example PIXIE frequency coverage spans 15 GHz to 3 THz in 15 GHz wide channels and has a per-channel sensitivity of approximately $5~\mathrm{Jy\sqrt{yr}~sr^{-1}}$.\footnote{PIXIE also includes channels beyond 3 THz, but these are not useful for our purposes.} The sensitivity was derived in~\cite{Kogut2011PIXIE} assuming photon noise limited detectors and a $4~\mathrm{cm^2~sr}$ etendu. The photon noise is dominated by the CMB, thermal dust, and zodiacal emission. Enhanced versions of PIXIE, using multiple FTS copies, have been considered more recently \citep{PRISM2013WPII, Kogut2019BAAS}. The primary way to improve the sensitivity for an FTS is to increase the etendu (not the number of detectors as is typical with imaging experiments), thereby increasing the number of photons collected and improving the signal-to-noise by a factor of the square root of the etendu. There is also a trade-off between the FTS frequency resolution and sensitivity per channel. Given a fixed optical loading and integration time, increasing the frequency resolution by a factor $m$ would decrease the sensitivity per channel linearly by $m$.~\footnote{The inverse relation between frequency resolution and sensitivity per channel means one loses a factor of $\sqrt{m}$ when averaging channels post-hoc. Thus, one cannot take observations with a very fine frequency resolution and average them together in the analysis without paying a penalty in signal-to-noise.}

\begin{deluxetable*}{l l l l l}[ht]
\caption{Uncertainty on $H_0$ for the two example noise curves with different modeling assumptions. The first column describes the modeling scenario, the second column gives the resulting effective sensitivity, the third column shows the forecast for $\sigma_{H_0}$, and the last two columns list the detection significance in units of standard deviations and percent. We use a Fisher matrix to calculate $s_T$ given a noise curve and a parametric model for the signals, both as a function of frequency. The CMB and foreground signals are discussed in Section~\ref{sec:fgs} and the noise curves are discussed in Section~\ref{sec:sens}. The propagation from $s_T$ to $\sigma_{H_0}$ is given by Equation~\ref{eqn:sigmaH0}. The cases without CMB spectral distortions and without foregrounds are shown to emphasize how dramatically the foregrounds degrade the results. The forecast labeled ``time-independent'' uses a noise curve integrated over 10 years to represent the potential gain from incorporating time-independent constraints. This gives a best-case forecast and would result in a better than $1\%$ measurement of $H_0$ with both noise curves. See Section~\ref{sec:time} for the discussion and motivation of the time-independent results. In all cases the total mission duration is 10 years (and note $s_T$ is listed in pico-Kelvin $\sqrt{\mathrm{yr}}$). \label{tab:sensitivity} }
\tablehead{\colhead{Scenario} & \colhead{$s_T~[\mathrm{pK\sqrt{yr}}]$} & \colhead{$\sigma_{H_0}~[\mathrm{km~s^{-1}~Mpc^{-1}}]$ } & \colhead{$X=H_0/\sigma_{H_0}$} & \colhead{$\sigma_{H_0}/H_0 (\%)$} }
\startdata
Linearly spaced noise curve & $54$ & $2.1$ & 33 & 3.0\% \\
... without foregrounds & $0.35$ & $0.014$ & 5100 & 0.020\% \\
... without CMB distortions & $1.3$ & $0.050$ & 1400 & 0.071\% \\ 
... time-independent & $17$ & $0.66$ & 110 & 0.94\% \\
\hline
Log spaced noise curve & $54$ & $2.1$ & 33 & 3.0\% \\
... without foregrounds & $0.88$ & $0.034$ & 2100 & 0.049\% \\
... without CMB distortions & $3.0$ & $0.12$ & 600 & 0.17\% \\ 
... time-independent & $17$ & $0.66$ & 110 & 0.94\% \\
\enddata
\end{deluxetable*}

The PIXIE sensitivity is not high enough to detect $H_0$, so we increase the sensitivity to produce two example noise curves that would measure $H_0$ to $3\%$, shown in Figure~\ref{fig:signals}. To achieve this we generate two example noise curves with broad frequency coverage and fine resolution and then set the sensitivity to realize a 3\% measurement of $H_0$. The example sensitivities are meant to serve as a guideline for what a future experiment might need to measure $H_0$. The first example has 2995 1~GHz wide channels linearly spaced between 5~GHz and 3~THz. The sensitivity per channel is 1.0~$\mathrm{mJy\sqrt{yr}~sr^{-1}}$ at low frequencies and turns up due to photon noise and low-pass filters at high frequencies (following the PIXIE noise behavior). The second example sensitivity curve has 1000 channels log-spaced between 5~GHz and 3~THz. The channel width begins at 30~MHz and increases to 20~GHz with a sensitivity of $7.5~\mathrm{mJy\sqrt{yr}~sr^{-1}}$ per channel (again following the shape derived from the PIXIE noise). For comparison, PIXIE and other recently proposed spectral distortion missions reach sensitivities between 0.1 and 10~$\mathrm{Jy\sqrt{yr}~sr^{-1}}$ per channel, depending on the specific configuration~\citep{Kogut2011PIXIE,Kogut2019BAAS,voyage2050}.

It is also worth noting that the raw instrument sensitivity of these examples is in the same range as recently proposed CMB imaging missions, such as the NASA-proposed Probe of Inflation and Cosmic Origins (PICO) satellite~\citep{PICOpaper}. The overall sensitivity after combining all of PICO's 21 frequency channels (noise-only) is $\approx0.5-1~\mathrm{\mu K}$-arcmin, obtained in a five-year mission. Averaging this over the full sky corresponds to a monopole sensitivity of $0.04 - 0.08~\mathrm{nK}$, i.e., $0.09 - 0.18~\mathrm{nK} \sqrt{\mathrm{yr}}$.  Evaluated at 150 GHz, this corresponds to $0.02-0.04~\mathrm{Jy~sr^{-1}}$, i.e., $0.04-0.09~\mathrm{Jy \sqrt{yr}~sr^{-1}}$, only a factor of a few larger than the per-channel sensitivities suggested here. While it is true that this exercise compares the per-channel sensitivity of the hypothetical mission here to the co-added sensitivity of all PICO channels, we note that the most sensitive PICO channels are only a factor of $\approx 2 - 3$ less sensitive than the co-added value used here.

%%%%%%%%%%%%%%%%%%%%%%%%%%%%%%%%%%%%%%%%%%%%%%%%%%%%%%%%%%%%%%%%%%%%%%%%%%%%%%%%%%%%%%%%
\subsection{Forecasts}
The Fisher forecast allows us to calculate the uncertainty on the CMB monopole temperature $T$ after one year of integration given the sky model and sensitivity. We assume the example mission evaluates the CMB temperature by analyzing the foreground and CMB monopole sky signal each year for 10 years. We then use Equation~\ref{eqn:sigmaH0} to propagate the uncertainty from $T$ to $H_0$. This assumes the signals are all time-dependent and have to be fit for in each year of integration, which is discussed in more detail in the next section.

Table~\ref{tab:sensitivity} presents the results of these forecasts. We include forecasts without CMB spectral distortions and without foregrounds to highlight that the foregrounds, which degrade the sensitivity to $H_0$ by several orders of magnitude, are the dominant hurdle for this type of measurement. The choice of bandwidth and frequency resolution of the experiment is a critical one, which depends on the signals and assumed foreground complexity. The log-spaced noise curve attempts to address this by placing more channels at low- and intermediate- frequencies ($\lesssim 300$~GHz) than the linearly spaced channel configuration that is natural to an FTS. This can be seen by comparing the nominal and without-foreground cases of Table~\ref{tab:sensitivity}. Both sensitivity curves measure $H_0$ to $3\%$ (as designed) in the nominal case. The degradation from foregrounds is a factor of $\approx 60$ for the log-spaced sensitivity and a factor of $\approx 150$ for the linearly-spaced sensitivity, meaning the log-spaced sensitivity is signifanctly more robust to foregrounds. In principle the frequency resolution and sensitivity curve could be optimized given the known CMB spectral distortion and foreground signals to minimize variance and bias; we leave this to future work. 

%%%%%%%%%%%%%%%%%%%%%%%%%%%%%%%%%%%%%%%%%%%%%%%%%%%%%%%%%%%%%%%%%%%%%%%%%%%%%%%%%%%%%%%%
\subsection{Time-independent signals}
\label{sec:time}
The Fisher forecast assumed that all the signals have to be fit for each year and that they are uncorrelated between years. This is not true in that all the CMB signals are being redshifted by the same amount and some of the foreground signals will not exhibit detectable time-variability on such short time scales, while others likely will. Assuming an ideal experiment with excellent stability over the mission would allow for the time-dependence of the CMB spectral distortions to be explicitly modeled with fewer parameters than the fully time-dependent version. This would improve the resulting constraint on $H_0$. Additionally, the foreground time variations could be parameterized as well (e.g., with a polynomial in time) and again reduce the dimensionality of the problem. However, in practice time-dependent systematic effects might reduce some of these gains.

The Galactic foregrounds exhibit intrinsic time variability from physical processes in the evolution of the Galaxy. These features should be small over tens of years, but given that the dynamical timescale of the Galaxy is on the order of 10-100 Myr, we might expect changes in the foreground intensity on order of $\frac{\mathrm{mission\ duration}}{\mathrm{Gal.\ time\ scale}}\approx 10^{-6}-10^{-7}$, comparable to the size of the signal of interest. A similar argument can be made for the extra-Galactic foregrounds. The integrated CO and CIB will exhibit both intrinsic time evolution due to galaxy evolution as well as extrinsic redshifting with the expansion of space. The redshifting of the extra-Galactic signals will be the same as the CMB and could therefore be modeled in the same way, potentially even improving the $H_0$ constraint.

Ideally we would calculate the uncertainties from a full frequency-time model with years $\times$ number of parameters per year $\approx120$ total parameters. The inversion of the Fisher matrix turns out to be numerically difficult in this case, as the signals are nearly degenerate for small variations in the parameters. Instead, we forecast the time-independent gains by bounding the forecast from above and below. The fully time-dependent forecast that we presented previously serves as the upper bound. For the lower bound we use a 10 year sensitivity in the forecasts, which represents the assumption that all of the signals are time-independent over the mission's duration. The constraint on $H_0$ in the time-independent case is then approximately a factor of three better than the time-dependent case. The full time-frequency model would give a result between these bounds, as shown in Table~\ref{tab:sensitivity}.

%%%%%%%%%%%%%%%%%%%%%%%%%%%%%%%%%%%%%%%%%%%%%%%%%%%%%%%%%%%%%%%%%%%%%%%%%%%%%%%%%%%%%%%%
\section{Discussion}
\label{sec:discussion}
Measuring the expansion of space from the cooling of the CMB benefits from being theoretically direct and requiring few cosmological assumptions as suggested by~\citet{zibin2007cmbevolution}. However it suffers from a variety of practical difficulties, including marginalization over foreground signals and requiring high sensitivity, fine frequency resolution, long-term stability, and precise calibration. The target sensitivity at this time seems out of reach given the requisite spectral sensitivity and resolution, without significant investment. The suggested log-spaced channels in principle are possible with an FTS with a time-dependent modulated mirror speed, but the sensitivity per channel would no longer be uniform in this case. In this sense the target sensitivity might be difficult to achieve using a single FTS but could potentially be approximated by multiple complementary FTS instruments, such as the instrument discussed for Super-PIXIE~\citep{Kogut2019BAAS}. An FTS is not necessarily the only approach, and a futuristic focal plane with many frequency channels (hundreds to thousands) might be more appropriate. Additionally, new foreground analysis methods and optimization of the sensitivity curve are under investigation and might alleviate some of the instrumental requirements. It should also be noted that an experiment capable of measuring $H_0$ from the cooling of the CMB would also detect CMB spectral distortions at high significance~\citep[e.g.,][]{Hill2015}, thereby testing our understanding of the early universe and particle physics \citep[e.g.,][]{voyage2050}, which provides further strong motivation. 

We find that this conclusion is primarily a foreground problem, as the foregrounds are what drive the strict instrument requirements. Without foregrounds an optimal sensitivity configuration would require far fewer channels and would be achievable with current technology. For example, modifying PIXIE to have 100~GHz wide channels would significantly improve the sensitivity per channel and allow for a $1\sigma$ measurement of $H_0$, without foregrounds. The calibration and long-term stability are then the primary experimental challenges.

A variety of additional complexities may arise other than the foregrounds and their variability when considering signals at this level. The experiment stability will be a primary concern, requiring sub-nK stability over 10 years. Additional contributions to the monopole might also have to be carefully considered. For example, the dipole and higher order multipoles will leak into the monopole given a partial sky average and light aberration \citep[][]{Chluba2011ab, Balashev2015, siavash2017mono}. The dipole itself changes due to our Galactic motion~\citep{moss2008cmbevolution} and the local gravitational potential from our Galaxy and local group contribute a Sachs-Wolfe like blueshifting to the monopole.
There are potentially other CMB spectral distortions such as the residual distortion that results from time-dependent energy injection in the early universe~\citep{Chluba2013PCA}. However, we are only interested in the time-evolution of the monopole and all these signals should not vary significantly over short time scales. Finally, the initial value of $T$ is not important and could even be biased to some degree without significantly affecting the constraint on $H_0$ from the slope of $T(t)$. Therefore these issues are likely to be subdominant to the foreground problem. 

%%%%%%%%%%%%%%%%%%%%%%%%%%%%%%%%%%%%%%%%%%%%%%%%%%%%%%%%%%%%%%%%%%%%%%%%%%%%%%%%%%%%%%%%
{\small
\section*{Acknowledgments}
We thank the referees for their helpful suggestions. The authors are grateful to David Alonso, Neal Dalal, Pedro Ferreira, and Aditya Rotti for helpful discussions. JC is supported by the Royal Society as a Royal Society University Research Fellow at the University of Manchester, UK, and the European Research Council (ERC) consolidator grant {\it CMBSPEC} (No.~725456). JCH acknowledges support from the W. M. Keck Foundation Fund at the Institute for Advanced Study, and from the Simons Foundation. MHA is supported by a Beecroft Fellowship at the University of Oxford. This project has received funding from the ERC under the European Union’s Horizon 2020 research and innovation programme (grant agreement No.~693024).}

%%%%%%%%%%%%%%%%%%%%%%%%%%%%%%%%%%%%%%%%%%%%%%%%%%%%%%%%%%%%%%%%%%%%%%%%%%%%%%%%%%%%%%%%
\bibliography{Lit}
\end{document}